\documentclass[conference]{IEEEtran}
\IEEEoverridecommandlockouts
\def\BibTeX{{\rm B\kern-.05em{\sc i\kern-.025em b}\kern-.08em
		T\kern-.1667em\lower.7ex\hbox{E}\kern-.125emX}}
	
\usepackage{balance}
\usepackage{cite}
\usepackage{amsmath,amssymb,amsfonts}
\usepackage{algorithmic}
\usepackage{graphicx}
\usepackage{textcomp}
\usepackage{xcolor}
\usepackage{todonotes}
\usepackage{booktabs}
\usepackage{multicol}

\usepackage{url}

\begin{document}

\title{Towards Systematic Specification and Verification of Fairness Requirements: A Position Paper}

\author{
	\IEEEauthorblockN{Qusai Ramadan\IEEEauthorrefmark{1},  
Jukka Ruohonen\IEEEauthorrefmark{1},  Abhishek Tiwari\IEEEauthorrefmark{1}, Adam Alami\IEEEauthorrefmark{1}, Zeyd Boukhers\IEEEauthorrefmark{2}
    }
\IEEEauthorblockA{\IEEEauthorrefmark{1}
University of Southern Denmark \\ \{qura, juk, abti, adal\}@mmmi.sdu.dk}

\IEEEauthorblockA{\IEEEauthorrefmark{2}Fraunhofer FIT, Germany \\ zeyd.boukhers@fit.fraunhofer.de}

}

\maketitle

\begin{abstract}
	Decisions suggested by improperly designed software systems might be prone to discriminate against people based on protected characteristics, such as gender and ethnicity. 
	Previous studies attribute such undesired behavior to flaws in algorithmic design or biased data.
	However, these studies ignore that discrimination is often the result of a lack of well-specified fairness requirements and their verification.
	The fact that experts' knowledge about fairness is often implicit makes the task of specifying precise and verifiable fairness requirements difficult. 
	In related domains, such as security engineering, knowledge graphs have been proven to be effective in formalizing knowledge to assist requirements specification and verification. 
	To address the lack of formal mechanisms for specifying and verifying fairness requirements, we propose the development of a knowledge graph-based framework for fairness. In this paper, we discuss the challenges, research questions, and a road map towards addressing the research questions.	


\end{abstract}

\section{Introduction}
\label{sec:intro}
Today's artificial intelligence (AI) systems provide predictions for sensitive decisions with far-reaching societal impact in many areas of our lives, such as who will be invited to a job interview, who will receive what kind of medical treatment, or whether a sentenced person should be kept in prison or can be released. Despite the benefits of such AI systems, there is a risk that  decisions suggested by AI systems with flawed design assumptions or insufficient fairness considerations are prone to discriminate against people based on protected characteristics such as gender and ethnicity.  For instance, in 2019, a study found racial bias in a widely used health care algorithm by many hospitals in the United States to predict who needs extra medical care \cite{Obermeyer19}.  Specifically, the hospital system systematically discriminated against patients of African descent by using health care costs as a proxy for medical need, which led to underestimating the severity of illness compared to equally sick patients of European descent.

According to Saxena, et al. \cite{saxena2020fairness}, \textit{fairness can be defined as the absence of any prejudice or
favoritism toward an individual or a group based on their
inherent or acquired characteristics}. However, this definition is open to a variety of interpretations. Philosophers have been debating for centuries on how to define and measure fairness \cite{rawls1991justice}. Similarly, psychologists have dedicated decades to exploring how people perceive fairness \cite{jetten2019social}. 
Nowadays, a diverse community of scholars is aware of the impact of discrimination and has begun researching the development of fair AI-based decision-making systems to avoid algorithmic bias. For example, the ACM Conference on Fairness, Accountability, and Transparency (FAccT)\footnote{https://facctconference.org/} brought together computer scientists, lawyers, and social scientists to formalize the vague notions of fairness in society to incorporate fairness in the AI system development. Thus far, many formal measures of fairness have emerged, including (e.g., demographic parity, equal opportunity, individual fairness, etc) \cite{Arvind18}. But while the different measures of fairness are \textit{contextual}, \textit{procedural}, \textit{contestable}, and \textit{hard to be achieved simultaneously} \cite{selbst2019fairness,brun2018software}, existing approaches for analyzing software fairness (except the work in \cite{ramadan2025mbfair} which is discussed in the state-of-the-art) focus on \textit{ex-post} analysis, either during the testing phase (e.g., \cite{galhotra2017fairness,datta2017proxy}) or at run-time (e.g., \cite{albarghouthi2017fairsquare,albarghouthi2019fairness}). These methods rely on hard-coded constraints and rigid formal definitions, limiting their ability to address fairness issues proactively. This narrow focus fundamentally ignores the root cause of discrimination due to the absence of a systematically designed fairness-aware system. According to Brun et al. \cite{brun2018software} ``\textit{as with software security [...], fairness needs to be a first-class entity in the software engineering process}''.

\vspace{1ex}
\noindent\textbf{Objectives.} As a society, our collective responsibility is to ensure fairness in AI systems by avoiding the root source of potential discrimination right from the onset of system development. To achieve this goal we aim at realizing the following objectives: 

\begin{itemize}
    \item \textbf{Objective 1}. To empower requirements engineers with an approach that enable them in formulating \textit{unambiguous} and \textit{verifiable} fairness requirements.

    \item \textbf{Objective 2}. To develop an approach that \textit{systematically} verifies compliance with fairness requirements during the design and implementation phases of the system.
\end{itemize}

Realizing these objectives is a challenging task. In this paper, we highlight the research challenges and questions, and provide a road map toward addressing them.  This paper is organized as follows. Section \ref{sec:back} provides the necessary background. Section \ref{sec:ChRe} discuss the challenges and research questions (RQs).  Section \ref{sec:Method}  discusses our ongoing work towards a systematic specification and verification of fairness requirements. Finally, Section \ref{sec:conclusion} concludes our key arguments for the proposed line of research.

\section{Background}
\label{sec:back}

In what follows, we elaborate the fundamental concepts of software fairness

\smallskip
\noindent\textit{Protected characteristics} refer to data that should not commonly be used to differentiate between people in a decision-making process \cite{verma2018fairness} such as gender, ethnicity, or age. Note that the concept is broader than the concept of sensitive personal data, as defined in Article 9 of the General Data Protection Regulation (GDPR) of the European Union (EU). Furthermore, to put aside national laws, there are two EU laws that address non-discrimination explicitly; in general, according to the laws, no one should be discriminated based on protected characteristics~\cite{Ruohonen24}. Furthermore, the EU's new AI Act obliges to audit high-risk AI systems for biases that may lead to discrimination. Thus, the proposal is relevant also from a legal perspective.

\smallskip
\noindent\textit{Proxy information} refers to data that are highly correlated with protected characteristics \cite{datta2017proxy,albarghouthi2019fairness, Ruohonen24}. For example, if young people are more likely to be healthy than older people, then the health status is a proxy for age. Proxy information can be identified manually by domain experts or automatically using a database of personal information.

\smallskip
\noindent\textit{Explanatory data attributes} refer to data attributes for which reasonable justification exists to be used in a given decision-making software even if that leads to discrimination on the basis of protected characteristics or proxy information \cite{tramer2015discovering}. 

%

\smallskip
\noindent\textit{Fairness Measures}
There have emerged several fairness measures in the literature \cite{Arvind18}. Examples of distinguished fairness measures include \textit{individual fairness} and \textit{group fairness}.
A decision-making software preserves individual fairness if it produces the same decision for every two individuals whose input data is identical except for the protected characteristics~\cite{dwork2012fairness}. 
Simultaneously, a software preserves group fairness if it produces equally distributed outputs for each protected group \cite{verma2018fairness}. However, prioritizing group fairness may be perceived as unfair under other fairness criteria, such as individual fairness. This highlights the broader challenge in algorithmic fairness. The fairness of a system often depends on the specific notion of fairness adopted, which is itself open to ethical and contextual interpretation.

\section{Challenges and Research Questions}
\label{sec:ChRe}

Realizing the two objectives stated in the introduction is a challenging task. In this section, we explain the challenges and propose open research questions.

\vspace{1ex}
\noindent\textbf{Challenge 1.} Specifying unambiguous and verifiable fairness requirements (Objective 1) requires awareness of several fairness aspects including data subject to be protected, fairness notion to be considered (e.g., demographic parity, equal opportunity, predictive parity,  individual fairness, etc.), protected characteristics and their proxies in specific context. Yet, it is unclear which fairness aspects have to be covered when writing fairness requirements~\cite{brun2018software}. One key factor contributes to this challenge is that domain and legal knowledge about fairness are often implicit and distributed across many documents. For example, consider the following requirement for a loan approval system:

\begin{quote}
    \textit{``The loan approval system must ensure that applicants are treated fairly regardless of their financial background.''}
\end{quote}

This requirement is vague and hard to verify. It does not specify what ``treated fairly'' means—whether in terms of approval rates, false positives, or other metrics. The phrase ``financial background'' is also unclear and may include attributes like income, employment history, or residence, which can serve as proxies for protected characteristics. More critically, it ignores domain-specific constraints. In lending, fairness is shaped not only by ethical considerations but also by legal obligations, such as those defined in the USA Equal Credit Opportunity Act \footnote{\url{https://www.ftc.gov/legal-library/browse/statutes/equal-credit-opportunity-act}}. Without aligning the requirement with such domain-specific rules, it remains too abstract for implementation, verification or compliance. 

Based on the context and legal policies, a well-specified fairness requirement should be certain about several aspects including: \begin{enumerate}
    \item The data subjects to be protected.
    \item The protected characteristics that should not commonly be used in that context to differentiate between data subjects.
    \item The data attributes whose use can be reasonably justified in that context, even if such use results in discrimination based on protected characteristics. For example, it is permitted to disallow blind applicants to take a driving license test.
    \item The notion of fairness that should be ensured.
\end{enumerate}

Failure to account for the aspects mentioned above and many other dynamic system aspects may lead inexpert developers to fall into the \textit{formalism trap}~\cite{selbst2019fairness}. For example, in 2016, journalists at ProPublica demonstrated that the COMPAS system for predicting re-offending risk score violates a certain measure of fairness, namely, equality of error rates \cite{ProPublica}. However, the founder of COMPAS (Northpointe) argued that it was fair because the system ensured other fairness measures, namely, accuracy equality and predictive parity \cite{dieterich2016compas}. Research question RQ1 aims at addressing this challenge: 

\begin{itemize}
    \item \textbf{RQ1}. How can we assist requirements engineers in specifying unambiguous and verifiable fairness requirements?
\end{itemize}

\vspace{2ex}
\noindent\textbf{Challenge 2.} Ensuring compliance with fairness requirements throughout the design and implementation of AI software (Objective 2) is challenging.  Only avoiding protected characteristics in decision-making software does not prevent discrimination. Due to data correlations, other data may act as proxies for protected characteristics, thereby causing \textit{indirect discrimination}~\cite{galhotra2017fairness}.  For example, in 2016, a delivery decision-making software by Amazon excluded some neighborhoods with African American communities in the United States from being able to participate in a free-delivery service, although the software did not explicitly use the ethnicity of the customers for making the decisions. 

In general, there are two main explanations for data correlations \cite{ramadan2018model}: 
(i) \textit{Societal fact}: For instance, if females are more likely to have long
hair more than males, then long hair can act as a proxy for the gender. (ii) \textit{data flow}: The actual input of a decision-making software may contain data resulting from processed protected characteristics.  The latter issue is typical for the problem that Dwork et al. \cite{dwork2018individual} observed: \textit{"fairness is not a property of software in isolation, fairness is a property of the software composed into its system"}.  Consider, for example, a common pipeline in modern banking systems involving multiple AI components. The first component handles identity verification and anti-money laundering (AML) risk assessment. It processes sensitive data such as nationality, place of birth, and residency, and outputs a risk score. In this context, the use of protected characteristics is legally justified due to regulatory obligations under Know Your Customer (KYC) and AML requirements. However, this risk score is then passed to a second AI component responsible for assigning credit lines to new customers. In this context, the credit decision model can exhibit biased behavior due to indirect information flow. Whether this is legally permitted or not, our role as software engineers is to uncover and address any potential discrimination in the system before it occurs and ensure it is properly reported. The second research question addresses this challenge:

\begin{itemize}
    \item  \textbf{RQ2}. How can fairness requirements be systematically enforced throughout the modeling and implementation of AI systems?
\end{itemize}

The RQ2's relevance is reinforced by the third challenge:

\vspace{1ex}
\noindent\textbf{Challenge 3.} According to the EU's laws, discrimination is prohibited and high-risk AI systems should be audited against discrimination and other biases, but, at the same time, the GDPR generally discourages and in most cases strictly prohibits the collection and processing of sensitive personal data~\cite{Ruohonen24}. This legal conundrum makes the enforcement envisioned in RQ2 even more difficult to achieve on one hand and relevant on the other hand.
\section{State Of The Art}
\label{sec:SOTA}

Related work can be classified into three streams of work. First, approaches to understanding fairness focus on defining fairness, both formally and informally \cite{verma2018fairness,ntoutsi2020bias,hutchinson201950}. Second, approaches to mitigating discrimination aim to prevent unfair treatment in AI software \cite{calmon2017optimized,zafar2017fairness,kamiran2018exploiting}. Third, approaches for discrimination detection focus on testing and verifying whether a system is fair with respect to a specific fairness notion \cite{adebayo2016fairml,tramer2015discovering,galhotra2017fairness}. \textit{Despite the many existing approaches in software fairness, no research supports the systematic specification and verification of fairness requirements}. In the following, we discuss the most relevant approaches to this paper.

\smallskip
\noindent\textbf{Requirements engineering approaches.}  The authors in \cite{verma2018fairness, baresi2023understanding}  proposed taxonomies to help the expert
decide and choose the right interpretation of fairness. However, none of these are context aware or enable  specifying verifiable fairness requirements.  In \cite{Farahani21}, the authors provide a goal-oriented approach for eliciting fairness requirements while taking into consideration the dynamic aspects of the system in question. However, as also stated in \cite{Farahani21}, the requirements resulting from this method may be subjective and uncertain as they remain on the goal level.  Hence, refining the fairness goals into more operational form is still an open problem.

\smallskip
\noindent\textbf{Model-based engineering approaches.} The need to integrate fairness as a first-class entity in the early stages of the software development process has been motivated in \cite{ramadan2018model} and \cite{brun2018software}. However, these works do not provide any technical approach that realizes their motivation.  In \cite{ramadan2020semi} the authors proposed a BPMN-based framework that supports the detection of conflicts between data protection-related goals, including fairness. However, the approach deals with fairness at the goal level, rather than assisting in their specification \cite{ramadan2020semi}.

\begin{figure*}[t!]
	\center
\includegraphics[width=\textwidth]{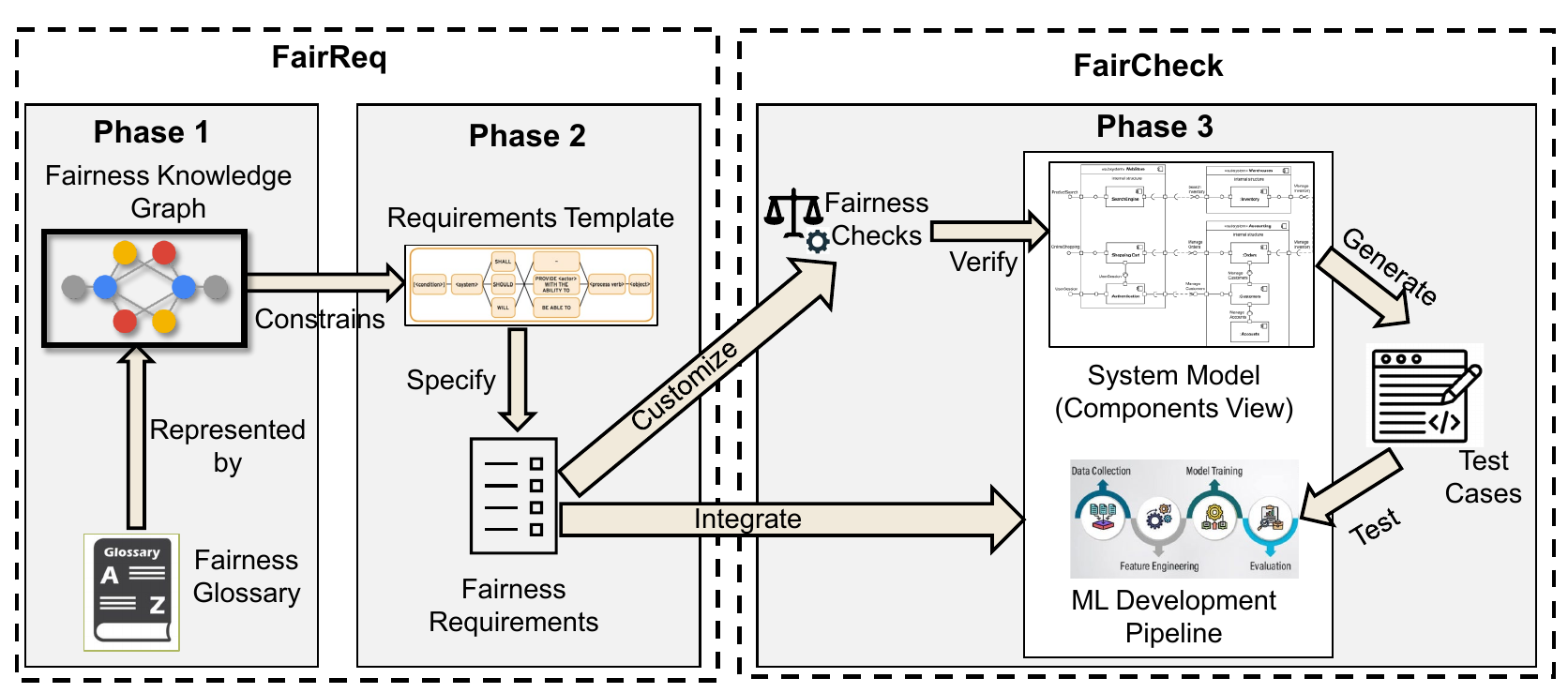}
	\caption{Conceptual and simplified architecture for our  proposed methodology}
	\label{fig:fairReq}
\end{figure*}

Furthermore, in \cite{ramadan2025mbfair}, the authors developed a model-based methodology called MBFair for verifying UML-based software designs with respect to individual fairness. MBFair performs verification by generating temporal logic clauses and analyzing their results. However, MBFair is limited to if-then-else decision-making software and does not support machine learning-based systems. Additionally, it only verifies individual fairness, without considering other statistical notions of fairness such as Demographic Parity, Equal Opportunity, Equalized Odds and many others \cite{Arvind18}.

\smallskip
\noindent\textbf{Fairness of Decision-Making AI Systems.} In addition to fairness-aware requirements engineering and model-based design, a significant body of work focuses on evaluating fairness in AI systems \textit{after deployment or implementation}. These approaches aim to uncover discriminatory behavior in decision-making software by analyzing its inputs/outputs. For example, FairTest~\cite{tramer2015discovering} and Themis~\cite{galhotra2017fairness} are two methods that uncover discrimination in a software system by studying the correlations between its outputs and inputs.
Different from FairTest~\cite{tramer2015discovering}, which requires a data set of possible inputs, Themis \cite{galhotra2017fairness} automatically generates input test cases based on a schema describing valid software inputs.

The authors of \cite{ruggieri2010data} proposed a method that extracts classification rules from a dataset of historical decision records. The classification rules are then directly mined to search for discrimination with respect to protected characteristics and their proxies. In \cite{zhang2016causal}, the authors propose an algorithm for detecting discrimination by analyzing a causal network that captures the causal structure among the data stored in historical decision records.   
In \cite{datta2015automated}, a tool called AdFisher is proposed. AdFisher aims to detect violations of individual fairness by monitoring the changes in the ads that users can receive based on web users' behaviors and privacy preferences specified in their web browsers.

In \cite{datta2017proxy}, the authors proposed a \textit{white-box} approach that analyzes decision-making software and returns witnesses that describe how  discrimination happened. FairSquare \cite{ADDN2017} is another white-box approach that is concerned with detecting discrimination \textit{at runtime}. 
FairSquare verifies probabilistic assertions in the source code of decision-making AI system software during its execution.

The approaches discussed above can be used as fairness testing components during the \textit{testing phase} of the system life cycle or at \textit{runtime}, which is only after the software is implemented. All of them apply hard-coded constraints to allow evaluating the system concerning a specific formalized definition of fairness. 
However, none of them provides guidelines for when each definition should be used or whether, for example, a decision that is based on proxy information in a specific context is permitted.

\section{Roadmap: Developing A Systematic Framework For Specifying and Verifying Fairness Requirements}
\label{sec:Method}

\begin{figure}[h!]
	\center
	\includegraphics[width=\columnwidth]{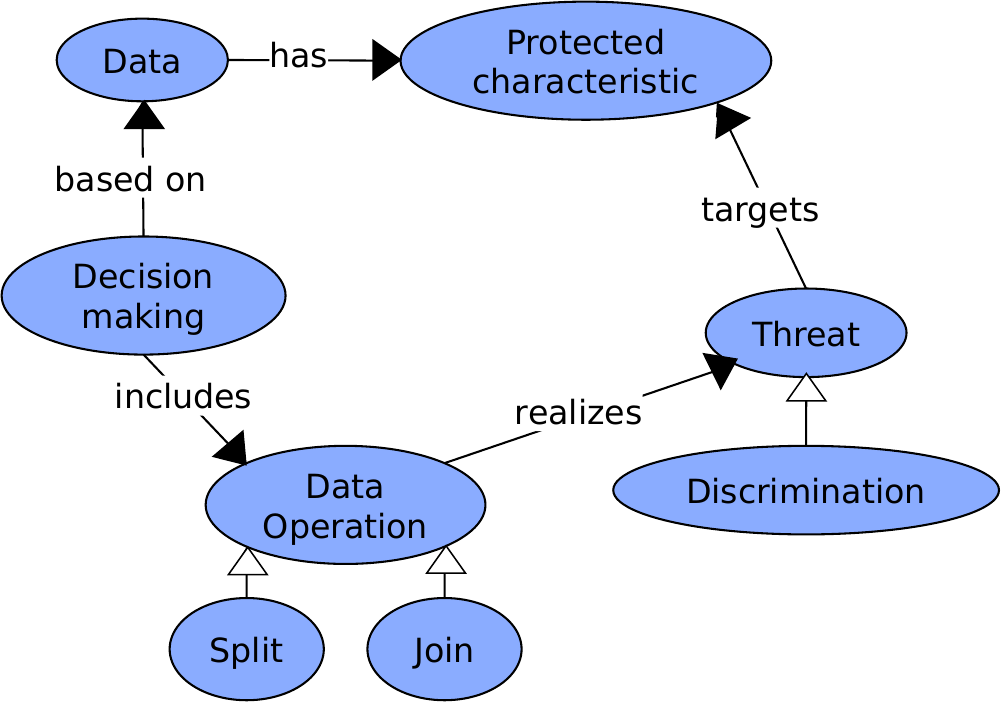}
	\caption{Simplified Example of fairness knowledge graph}
	\label{fig:graph}
\end{figure}

In critical domains knowledge graphs have been proven effective in assisting requirements specification and verification \cite{PBKJ2021,SSMC2005}. For example, in \cite{PBKJ2021}, Peldszus et al. use a layered ontology to represent security knowledge at the upper, domain, and system levels. When a vulnerability is discovered in an encryption algorithm, the knowledge graph evolves to reflect this change by introducing new concepts. This triggers automated updates in security requirements and checks to maintain system compliance with evolving security knowledge. Therefore, we hypothesize that: knowledge graphs can support the systematic specification and verification of fairness requirements throughout the design and implementation phases of AI systems. For this, we aim at developing a knowledge graph-based framework. Our framework consists of two main approaches, namely \textbf{FairReq} and \textbf{FairCheck}, which aim at addressing RQ1 and RQ2 respectively. Fig. \ref{fig:fairReq} provides an overview of the main components of our proposed framework. In the following, we present our planned roadmap toward developing the framework. Our framework consists of three development phases as follows:

\smallskip
\noindent \textbf{Phase 1. Fairness Knowledge Representation.} In this phase, we plan to: (1) Perform a systematic review to the state-of-the-art and relevant laws to establish a good understanding for different notions of fairness and their context. This task involves also the concept of legal requirements used in requirements engineering research. (2) Develop a machine-readable glossary using SKOS (Simple Knowledge Organization System). The glossary will contain different fairness-related concepts along with their definitions. (3) Develop a knowledge graph (e.g., using ontology language) that captures the relationships between different concepts in the glossary. The knowledge graph will enable engineers in practical requirements engineering to protect against the design of discriminatory AI systems. In order to illustrate a possible structure for such a graph, we provide a first simplified representation in Fig. \ref{fig:graph}. However, in our work we envision two interconnected knowledge graphs, where the first remains independent of any specific domain and captures general software fairness knowledge in addition to relevant legal constraints based on AI Act\footnote{https://eur-lex.europa.eu/eli/reg/2024/1689/oj/eng}. The second graph represents domain-specific knowledge. The following is a description of Fig. \ref{fig:graph}.

\textit{Data}: The starting point of an AI system is the underlying data. Depending on the application and the target of the decision, these can be available in different formats, e.g., structured, semi-structured, and unstructured. 

\textit{Protected Characteristics}: The provided data may explicitly contain protected characteristics for which decisions derived from an AI system are forbidden to discriminate against anyone based on these characteristics. Implicitly, discrimination may occur also through proxy variables, as was discussed.

\textit{Decision Making}: AI systems can utilize various approaches such as decision trees or deep neural networks for decision making. Thereby, all decision-making approaches can come with different fairness-related specifics.

\textit{Data Operation}: Depending on the current process phase, data operations relevant for an AI system's fairness could be divided into three categories: 1. Pre-processing, 2. In-processing, and 3. Post-processing. The detailed operations associated with these categories now depend heavily on the area or data of applications and the particular design of the AI system. Every data operation might realize a threat, e.g., joining protected characteristics with non-protected characteristics could lead to discrimination.

\textit{Threats}: In carrying out these operations on the data, various threats arise concerning fairness. For example, one of these is discrimination against a person based on protected characteristics. Here, the ontology should provide all possible threats and requirements engineers should be assisted in selecting the ones threatening the system they are designing. 

\textit{Fairness control}: To avoid fairness threats, corresponding fairness controls are necessary which finally then can be systematically derived from a semi-automated tool based on such an ontology.

\begin{figure}[t!]
	\center
	\includegraphics[width=\columnwidth]{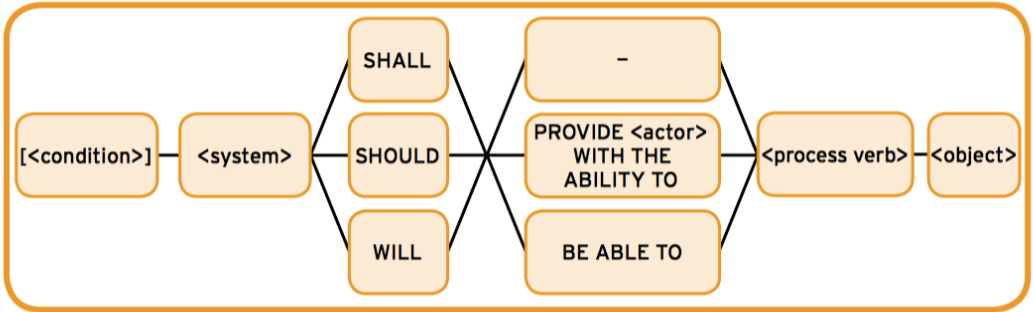}
	\caption{Functional MASTER template, extracted from \cite{Rupp2014b}}
	\label{fig:template}
\end{figure}

\begin{figure}[h!]
	\center
	\includegraphics[width=\columnwidth]{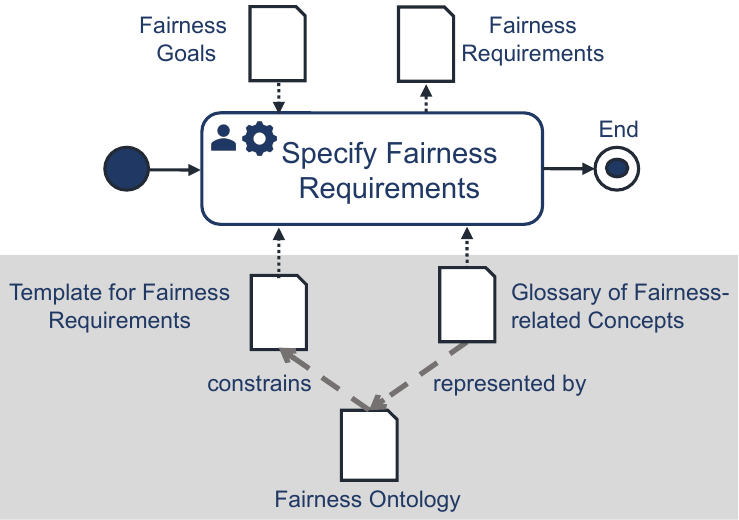}
	\caption{Specifying fairness requirements based on an ontology}
	\label{fig:working}
\end{figure}

\smallskip
\noindent \textbf{Phase 2. Fairness Requirements Specification.} In this phase we plan to: (1) Customize a requirement engineering template by extending the meta-model of a well-known template system (e.g.,  
 \textsc{MASTeR}, also known as \emph{Rupp templates} \cite{Rupp2014b}). For example, Fig. \ref{fig:template} shows the template used for specifying functional requirements in \textsc{MASTeR} template system. Generally, templates consist of fixed text and variable parts to be filled. Variable parts are often denoted within $<$~$>$ and optional parts with [ ]. By substituting the variable parts, a requirement is instantiated. (2) Develop a method to enable requirements engineering in specifying fairness requirements in a semi-automatic way. The interaction between customized template, glossary and knowledge graphs for such a semi-automated process can be then summarized as represented in Fig.~\ref{fig:working}.

\begin{figure}[t!]
	\center
	\includegraphics[width=\columnwidth]{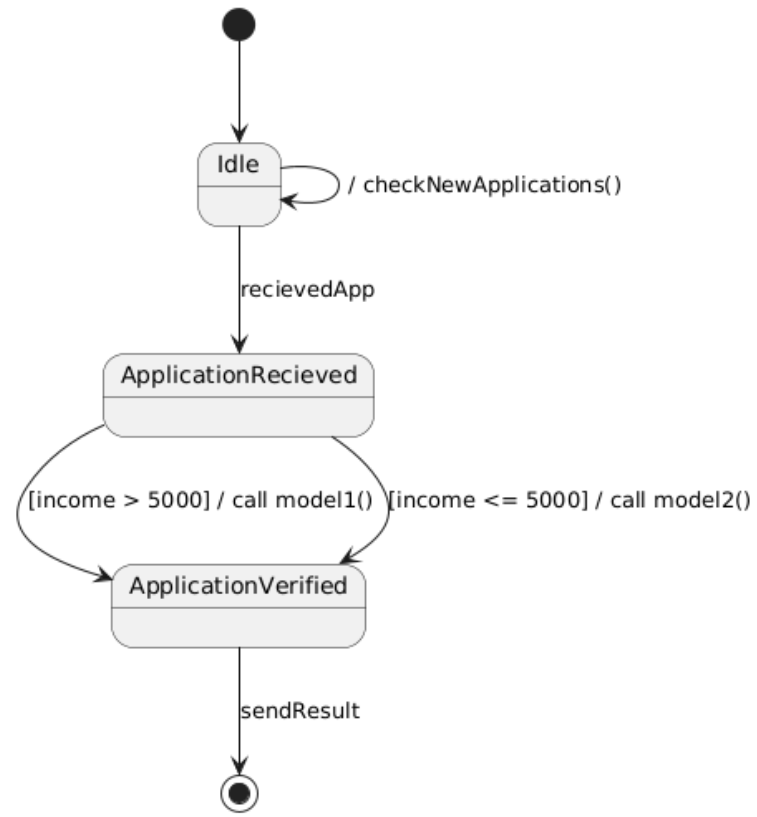}
	\caption{Example State Machine Diagram}
	\label{fig:stateMachine}
\end{figure}

\smallskip
\noindent \textbf{Phase 3. Integration and Verification.} In this phase we plan to: (1) Develop a method to automatically integrate the specified requirements into software models and source code by leveraging knowledge graph-driven transformations and rule-based mappings. This will involve generating model elements and annotated code structures based on formalized requirements. (2)  Develop system-level design-time checks. At this stage, we check fairness at the system architecture level while treating the ML component as a black box, meaning that no assumptions are made about its internal logic or output behavior. Instead, we assess the surrounding system structure and its data flows to detect fairness-relevant design flaws.  For example, consider Fig. \ref{fig:stateMachine} which illustrates a simplified UML state machine diagram for a loan processing system. Assume that \textit{income} is a proxy for \textit{age}. This assumption introduces a potential fairness concern early in the decision process. Specifically, the system starts in the \texttt{Idle} state and transitions to \texttt{ApplicationReceived} upon receiving a loan application. From there, a conditional decision is made based on income. If \texttt{income > 5000}, the system calls \texttt{model1()}. Otherwise,  it calls \texttt{model2()}. This design raises a potential risk of disparate treatment due to an indirect dependency on age. (3) Develop a method for generating abstract test cases using Model-Based Testing (MBT) tools like GraphWalker, ModelJUnit to confirm that protected attributes or their proxies do not influence flow or decisions.  While our analysis focuses on detecting fairness issues at the design level, we acknowledge that real-world deployments are constrained by privacy regulations such as the GDPR. These regulations (as explained earlier in challenge 3) restrict the collection and processing of sensitive personal data. To address this, we aim to generate synthetic data that reflects statistical patterns relevant for fairness evaluation without involving real individuals.

\section{Conclusion and Next Steps}
\label{sec:conclusion}

This paper advocates the need for an approach to support specifying precise and verifiable fairness requirements. Yet, there is still a limited common understanding of which fairness aspects should be addressed when specifying such requirements. We argue that failure to account for the context and certain fairness-specific aspects may lead to outcomes mis-aligned with our societal and human values, we all rely on.

To overcome this challenge, we propose to develop a knowledge graph-based framework for fairness to assist the specification of precise fairness requirements and enable their verification throughout the development.  To operationalize our framework, we plan to integrate it into CI/CD workflows as follows: Fairness requirements specified through structured templates would be compiled into machine-readable rules and linked to UML models via annotations. The machine-readable rules could then be propagated to the implementation as code annotations or configuration checks, verified through static analysis. In parallel, model-based test cases would be generated to validate fairness-related behavior. The tests would be executed automatically within the CI pipeline.

\bibliographystyle{IEEEtran}
\balance
\bibliography{literature}

\begin{thebibliography}{10}
\providecommand{\url}[1]{#1}
\csname url@samestyle\endcsname
\providecommand{\newblock}{\relax}
\providecommand{\bibinfo}[2]{#2}
\providecommand{\BIBentrySTDinterwordspacing}{\spaceskip=0pt\relax}
\providecommand{\BIBentryALTinterwordstretchfactor}{4}
\providecommand{\BIBentryALTinterwordspacing}{\spaceskip=\fontdimen2\font plus
\BIBentryALTinterwordstretchfactor\fontdimen3\font minus \fontdimen4\font\relax}
\providecommand{\BIBforeignlanguage}[2]{{%
\expandafter\ifx\csname l@#1\endcsname\relax
\typeout{** WARNING: IEEEtran.bst: No hyphenation pattern has been}%
\typeout{** loaded for the language `#1'. Using the pattern for}%
\typeout{** the default language instead.}%
\else
\language=\csname l@#1\endcsname
\fi
#2}}
\providecommand{\BIBdecl}{\relax}
\BIBdecl

\bibitem{Obermeyer19}
Z.~Obermeyer, B.~Powers, C.~Vogeli, and S.~Mullainathan, ``Dissecting racial bias in an algorithm used to manage the health of populations,'' \emph{Science}, vol. 366, no. 6464, 2019.

\bibitem{saxena2020fairness}
N.~A. Saxena, K.~Huang, E.~DeFilippis, G.~Radanovic, D.~C. Parkes, and Y.~Liu, ``How do fairness definitions fare? testing public attitudes towards three algorithmic definitions of fairness in loan allocations,'' \emph{Artificial Intelligence}, vol. 283, p. 103238, 2020.

\bibitem{rawls1991justice}
J.~Rawls, ``Justice as fairness: Political not metaphysical,'' in \emph{Equality and Liberty: Analyzing Rawls and Nozick}.\hskip 1em plus 0.5em minus 0.4em\relax Springer, 1991, pp. 145--173.

\bibitem{jetten2019social}
J.~Jetten and K.~Peters, \emph{The social psychology of inequality}.\hskip 1em plus 0.5em minus 0.4em\relax Springer, 2019.

\bibitem{Arvind18}
A.~Narayanan, ``21 fairness definitions and their politics,'' \emph{FAccT}, 2018.

\bibitem{selbst2019fairness}
A.~D. Selbst, D.~Boyd, S.~A. Friedler, S.~Venkatasubramanian, and J.~Vertesi, ``Fairness and abstraction in sociotechnical systems,'' in \emph{FAccT}, 2019.

\bibitem{brun2018software}
Y.~Brun and A.~Meliou, ``Software fairness,'' in \emph{ESEC/FSE}, 2018.

\bibitem{ramadan2025mbfair}
Q.~Ramadan, M.~Konersmann, A.~S. Ahmadian, J.~J{\"u}rjens, and S.~Staab, ``Mbfair: a model-based verification methodology for detecting violations of individual fairness,'' \emph{Software and Systems Modeling}, vol.~24, no.~1, pp. 111--136, 2025.

\bibitem{galhotra2017fairness}
S.~Galhotra, Y.~Brun, and A.~Meliou, ``Fairness testing: Testing software for discrimination,'' in \emph{FSE}, 2017.

\bibitem{datta2017proxy}
A.~Datta, M.~Fredrikson, G.~Ko, P.~Mardziel, and S.~Sen, ``Proxy non-discrimination in data-driven systems,'' \emph{arXiv preprint arXiv:1707.08120}, 2017.

\bibitem{albarghouthi2017fairsquare}
A.~Albarghouthi, L.~D'Antoni, S.~Drews, and A.~V. Nori, ``Fairsquare: probabilistic verification of program fairness,'' \emph{Proceedings of the ACM on Programming Languages}, vol.~1, no. OOPSLA, pp. 1--30, 2017.

\bibitem{albarghouthi2019fairness}
A.~Albarghouthi and S.~Vinitsky, ``Fairness-aware programming,'' in \emph{FAccT}, 2019.

\bibitem{verma2018fairness}
S.~Verma and J.~Rubin, ``Fairness definitions explained,'' in \emph{FairWare}, 2018.

\bibitem{Ruohonen24}
J.~Ruohonen, ``{O}n {A}lgorithmic {F}airness and the {EU} {R}egulations,'' 2024, {A}rchived manuscript, available online: \url{https://arxiv.org/abs/2411.08363}.

\bibitem{tramer2015discovering}
F.~Tram{\`e}r, V.~Atlidakis, R.~Geambasu, D.~J. Hsu, J.-P. Hubaux, M.~Humbert, A.~Juels, and H.~Lin, ``Discovering unwarranted associations in data-driven applications with the fairtest testing toolkit,'' \emph{CoRR}, 2015.

\bibitem{dwork2012fairness}
C.~Dwork, M.~Hardt, T.~Pitassi, O.~Reingold, and R.~Zemel, ``Fairness through awareness,'' in \emph{ITCS}, 2012, pp. 214--226.

\bibitem{ProPublica}
J.~Angwin, J.~Larson, S.~Mattu, and L.~Kirchner, ``Machine bias,'' \emph{ProPublica}, 2016.

\bibitem{dieterich2016compas}
W.~Dieterich, C.~Mendoza, and T.~Brennan, ``Compas risk scales: Demonstrating accuracy equity and predictive parity,'' \emph{Northpointe Inc}, vol.~7, no.~4, 2016.

\bibitem{ramadan2018model}
Q.~Ramadan, A.~S. Ahmadian, D.~Str{\"u}ber, J.~J{\"u}rjens, and S.~Staab, ``Model-based discrimination analysis: A position paper,'' in \emph{FairWare}, 2018.

\bibitem{dwork2018individual}
C.~Dwork and C.~Ilvento, ``Individual fairness under composition,'' \emph{Proceedings of fairness, accountability, transparency in machine learning}, 2018.

\bibitem{ntoutsi2020bias}
E.~Ntoutsi and o, ``Bias in data-driven artificial intelligence systems—an introductory survey,'' \emph{Wiley Interdisciplinary Reviews: Data Mining and Knowledge Discovery}, vol.~10, no.~3, 2020.

\bibitem{hutchinson201950}
B.~Hutchinson and M.~Mitchell, ``50 years of test (un) fairness: Lessons for machine learning,'' in \emph{FAccT}, 2019.

\bibitem{calmon2017optimized}
F.~Calmon, D.~Wei, B.~Vinzamuri, K.~N. Ramamurthy, and K.~R. Varshney, ``Optimized pre-processing for discrimination prevention,'' in \emph{NIPS}, 2017.

\bibitem{zafar2017fairness}
M.~B. Zafar, I.~Valera, M.~Gomez~Rodriguez, and K.~P. Gummadi, ``Fairness beyond disparate treatment \& disparate impact: Learning classification without disparate mistreatment,'' in \emph{WWW}, 2017.

\bibitem{kamiran2018exploiting}
F.~Kamiran \emph{et~al.}, ``Exploiting reject option in classification for social discrimination control,'' \emph{Inform Sciences}, vol. 425, 2018.

\bibitem{adebayo2016fairml}
J.~A. Adebayo \emph{et~al.}, ``Fairml: Toolbox for diagnosing bias in predictive modeling,'' Ph.D. dissertation, MIT, 2016.

\bibitem{baresi2023understanding}
L.~Baresi, C.~Criscuolo, and C.~Ghezzi, ``Understanding fairness requirements for ml-based software,'' in \emph{2023 IEEE 31st International Requirements Engineering Conference (RE)}.\hskip 1em plus 0.5em minus 0.4em\relax IEEE, 2023, pp. 341--346.

\bibitem{Farahani21}
A.~Farahani, L.~Pasquale, A.~Bennaceur, T.~Welsh, and B.~Nuseibeh, ``On adaptive fairness in software systems,'' in \emph{SEAMS}, 2021.

\bibitem{ramadan2020semi}
Q.~Ramadan, D.~Str{\"u}ber, M.~Salnitri, J.~J{\"u}rjens, V.~Riediger, and S.~Staab, ``A semi-automated bpmn-based framework for detecting conflicts between security, data-minimization, and fairness requirements,'' \emph{Softw Syst Model}, 2020.

\bibitem{ruggieri2010data}
S.~Ruggieri, D.~Pedreschi, and F.~Turini, ``Data mining for discrimination discovery,'' \emph{TKDD}, vol.~4, no.~2, p.~9, 2010.

\bibitem{zhang2016causal}
L.~Zhang, Y.~Wu, and X.~Wu, ``{On Discrimination Discovery Using Causal Networks},'' in \emph{International Conference on Social Computing, Behavioral-Cultural Modeling and Prediction and Behavior Representation in Modeling and Simulation}.\hskip 1em plus 0.5em minus 0.4em\relax Springer, 2016, pp. 83--93.

\bibitem{datta2015automated}
A.~Datta, M.~C. Tschantz, and A.~Datta, ``Automated experiments on ad privacy settings,'' in \emph{PoPETs}, 2015, pp. 92--112.

\bibitem{ADDN2017}
A.~Albarghouthi, L.~D'Antoni, S.~Drews, and A.~V. Nori, ``Fairsquare: Probabilistic verification of program fairness,'' in \emph{OOPSLA}, vol.~1, 2017.

\bibitem{PBKJ2021}
S.~Peldszus, J.~B{\"{u}}rger, T.~Kehrer, and J.~J{\"{u}}rjens, ``Ontology-driven evolution of software security,'' \emph{DKE}, vol. 134, 2021.

\bibitem{SSMC2005}
A.~Souag, C.~Salinesi, R.~Mazo, and I.~Comyn-Wattiau, ``A security ontology for security requirements elicitation,'' in \emph{ESSoS}, 2005.

\bibitem{Rupp2014b}
C.~Rupp and R.~Joppich, ``\BIBforeignlanguage{German}{{Anforderungsschablonen}},'' in \emph{\BIBforeignlanguage{German}{{R}equirements-{E}ngineering und -{M}anagement}}, 6th~ed.\hskip 1em plus 0.5em minus 0.4em\relax Carl Hanser Verlag M\"unchen, 2014, pp. 215--246.

\end{thebibliography}

\end{document}